# PALEONTOLOGICAL TESTS: HUMAN-LIKE INTELLIGENCE IS NOT A CONVERGENT FEATURE OF EVOLUTION

CHARLES H. LINEWEAVER
*Planetary Science Institute, Research School of Astronomy and Astrophysics, Research School of Earth Science,*
*Australian National University, Canberra, ACT 0200 Australia*

## 1. The Planet of the Apes Hypothesis

I taught a course called "Are We Alone?" at the University of New South Wales for a few years. The most popular lecture was "The Great Drake Equation Debate" – half a dozen "experts" would sit at the front of the crowded lecture theater defending their estimates for the various terms in the Drake Equation (an equation created by Frank Drake to estimate the number of civilizations in the Milky Way with whom we might communicate via radio telescopes). The first terms of the equation are astronomical. How many stars are in our galaxy? -- most experts agreed -- about 300 billion. What fraction of those stars are orbited by "Earth-like" planets? -- estimates ranged from ~100% to ~0.1% depending roughly proportionally on how specific "Earth-like" was interpreted to be. Then came the more contentious biological terms: What fraction of these Earth-like planets would harbor life? I defended a relatively high probability (~10%) based on how rapidly biogenesis occurred on Earth (Lineweaver & Davis 2002). We argued back and forth about how probable or improbable the steps of molecular evolution were, that led to life on Earth -- and whether there were places on Earth where life could still be emerging. We all learned a lot about biochemistry, autocatalytic cycles and hydrothermal vents. However, the most contentious term was: Once there is life of any kind, what is the probability that it will evolve into a human-like intelligence that can build and operate radio telescopes? (We define intelligence this way not out of some geeky technophilic perversity but because posed this way, we have the ability to answer the question by searching for other telescopes with our telescopes. So far, no signals from intelligent aliens have been identified, Tarter 2001).

In the "Great Drake Equation Debate" most of the invited experts assumed that once life got started it would get smarter and smarter until one day, it would hit upon the idea of building a radio telescope. Most students also subscribed to this "stupid things get smarter" model of animal evolution and believed it to be a universal trend. I call this idea the Planet of the Apes Hypothesis.

The movie *Planet of the Apes* (1968) is set in the future after a catastrophic nuclear war between *Homo sapiens*. The surviving humans have lost the ability to speak and have to forage in the wild. Meanwhile, three species of apes (chimps, gorillas and orangutans) learn to speak English, ride horses, farm corn, shoot rifles, and in general begin to act like hairy Victorian humanoids with human-like intelligence. They move into the recently emptied "intelligence niche" and turn into the "functional equivalent of humans" (to use Carl Sagan's term, Sagan 1995a).  On the Planet of the Apes, human-like intelligence is so adaptive that it is a convergent feature of evolution -- species are waiting in the wings to move in and occupy the intelligence niche.

G.G. Simpson in "The Nonprevalence of Humanoids" (1964) articulated the case that humans (or any given species) were a quirky product of terrestrial evolution and therefore we should not expect to find humanoids elsewhere.  Thus stupid things do not, in general acquire human-like intelligence.  The evidence we have tells us that once extinct, species do not re-evolve.  Evolution is irreversible.  This is known as Dollo's Law (Dollo 1893, Gould 1970).  The re-evolution of the same species is not something that happens only rarely.  It never has happened.  Simpson also argued that biologists, not physicists, can best judge this issue because the problem is one of evolutionary systematics, not deterministic physics.

Whether there is a trend in the fossil record indicating that stupid things tend to get smarter, is an important and controversial issue in which the discussion has become polarized into two camps. In one camp are the non-convergentists (mostly biologists) who, after studying the fossil record, insist that the series of events that led to human-like intelligence is not a trend, but a quirky result of events that will never repeat themselves either on Earth or anywhere else in the universe.  In the other camp are the convergentists (mostly physical scientists) who believe that stupid things get smarter and that intelligence is a convergent feature of evolution here and elsewhere.   See Lineweaver (2005) for more on the protagonists in this debate.

Is there any real evidence for the Planet of the Apes Hypothesis? Is human-like intelligence a convergent feature of evolution? Should we expect to find extraterrestrials with human-like intelligence?  Despite the lack of direct evidence, we would like to assemble and evaluate the best indirect evidence for or against the idea that life (terrestrial or extraterrestrial) evolves towards human-like intelligence.  The study of the evolutionary trends in the paleontological record of life on Earth is probably the most relevant evidence and that is what is critically examined here.

**2. Frank Drake, Carl Sagan and Ernst Mayr**

In 1960 Frank Drake conducted the first radio search for extraterrestrial intelligence.  He is the Director of the SETI (Search for Extra Terrestrial Intelligence) Institute's Center for the Study of Life in the Universe.  SETI work would seem much more promising if there has been an evolutionary trend among terrestrial life forms towards higher intelligence.

On a flight to a conference I asked him:  "Frank, why do you think there are intelligent aliens who have built radio telescopes?  What do you think is the strongest evidence for the idea that such human-like intelligence is a convergent feature of evolution?"  Frank's answer went something like this:

> "The Earth's fossil record is quite clear in showing that the complexity of the central nervous system - particularly the capabilities of the brain - has steadily increased in the course of evolution. Even the mass extinctions did not set back this steady increase in brain size. It can be argued that extinction events expedite the development of cognitive abilities, since those creatures with superior brains are better able to save themselves from the sudden change in their environment. Thus smarter creatures are selected, and the growth of intelligence accelerates. We see this effect in all varieties of animals -- it is not a fluke that has occurred in some small sub-set of animal life. This picture suggests strongly that, given enough time, a biota can evolve not just one intelligent species, but many. So complex life should occur abundantly." (Drake 2006)

During the flight Frank referred me to the debate between biologist Ernst Mayr and planetologist Carl Sagan (Sagan 1995a,b, Mayr 1995a,b). Sagan articulated the concept of "functionally equivalent humans":

"...when we're talking about extraterrestrial intelligence, we are not talking--despite Star Trek--of humans or humanoids. We are talking about the functional equivalent of humans-- say, any creatures able to build and operate radio telescopes." (Sagan 1995a)

"We are not requiring that they follow the particular route that led to the evolution of humans. There may be many different evolutionary pathways, each unlikely, but the sum of the number of pathways to intelligence may nevertheless be quite substantial." ( Sagan 1995a)

"...other things being equal, it is better to be smart than to be stupid, and an overall trend toward intelligence can be perceived in the fossil record. On some worlds, the selection pressure for intelligence may be higher; on others, lower." (Sagan 1995b)

To which Mayr replied:

"Sagan adopts the principle "it is better to be smart than to be stupid," but life on Earth refutes this claim. Among all the forms of life, neither the prokaryotes nor protists, fungi or plants has evolved smartness, as it should have if it were "better." In the 28 plus phyla of animals, intelligence evolved in only one (chordates) and doubtfully also in the cephalopods. And in the thousands of subdivisions of the chordates, high intelligence developed in only one, the primates, and even there only in one small subdivision. So much for the putative inevitability of the development of high intelligence because "it is better to be smart."( Mayr 1995b)

To which Sagan re-replied:

"Mayr argues that prokaryotes and protista have not "evolved smartness." Despite the great respect in which I hold Professor Mayr, I must demur: Prokaryotes and protista are our ancestors. They have evolved smartness, along with most of the rest of the gorgeous diversity of life on Earth." (Sagan 1995b).

Hold those poetic horses Carl! You're in Mayr's territory. By "prokaryotes and protists" Mayr is referring to extant organisms and their ancestors, **after** they diverged from our lineage. These ancestors and their lineages have continued to exist and evolve and have not produced intelligence. All together that makes about 3 billion years of prokaryotic evolution that did not produce high intelligence and about 600 million years of protist evolution that did not produce high intelligence.

### 3. The Fossil Evidence for an Overall Trend Towards Intelligence.

The comments by Drake (2006) and Sagan (1995b) about the fossil record showing evidence for a trend toward increasing vertebrate encephalization are references primarily to the work of paleoneurologist Harry Jerison (Jerison 1973, 1975, 1991). Jerison has been measuring the volume of modern and fossil animal skulls for several decades (Jerison 1975). Jerison introduced the concept of Encephalization Quotient (E.Q.), which is the ratio of brain weight to some power (~2/3 or ~3/4) of the body weight. E.Q. is arguably the most objective way to compare the intelligence of different groups of encephalated animals (Jerison 1955, 1963, 1973).

### 4. Interpretation Problem # 1: Selection Bias: Choosing E.Q. because we have the highest E.Q.

Every species has some unique feature — a feature that makes it different from its closest living relatives and from its ancestors. To make Fig. 1, Jerison first identified our unique feature (high E.Q.) and then plotted it as a function of time. Looking backward at the history of any existing extreme trait will produce a similar apparent trend. The evolutionary history of a feature (identified as extreme today) is almost guaranteed to look something like Fig. 1. Therefore, we cannot consider the trend seen in our lineage to be a general trend representative of any other lineages. If you choose a feature because of its current extreme nature, it is no surprise that it had to get that way and that its evolution will display a trend. But this trend has no claims to being representative of life in general.

Another example may make selection bias more obvious. Elephants have longer noses than their living relatives. Thus, when we focus on this unique feature and plot the sizes of the noses of its living relatives and of their evolutionary ancestors (Fig. 2), we find that in the series of progressively earlier ancestors, noses get progressively shorter. This is a selection effect that has nothing to do with a general tendency that can be extrapolated to the rest of biology. Increasing nose size is not a general feature of evolution. It is something that occurred in the lineage that led to elephants. After diverging from this lineage the N.Q. (nasalization quotient) of most groups stayed constant.

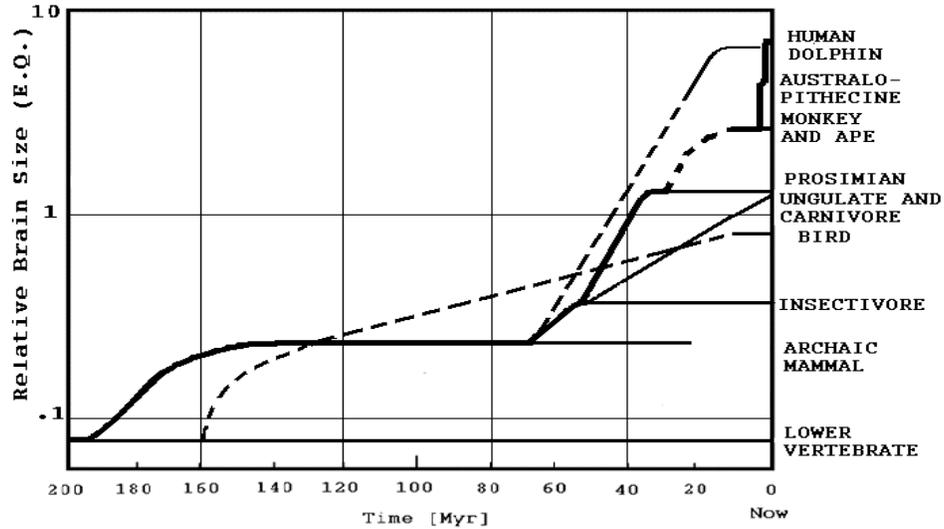

Figure 1. The Evolution of Relative Brain Size in Groups of Vertebrates Over the Past 200 Million Years (adapted and updated from Jerison 1976, p 96, Jerison 1991, Fig. 17). This plot seems to show an evolutionary trend towards increasing relative brain size ( = E.Q. = Encephalization Quotient) and is probably the most definitive evidence for such a trend. Average living mammal E.Q. is defined as 1. The broken lines indicate gaps in the fossil record. Variation within groups is not shown. The lineage that led to humans is drawn thicker than the other lineages.

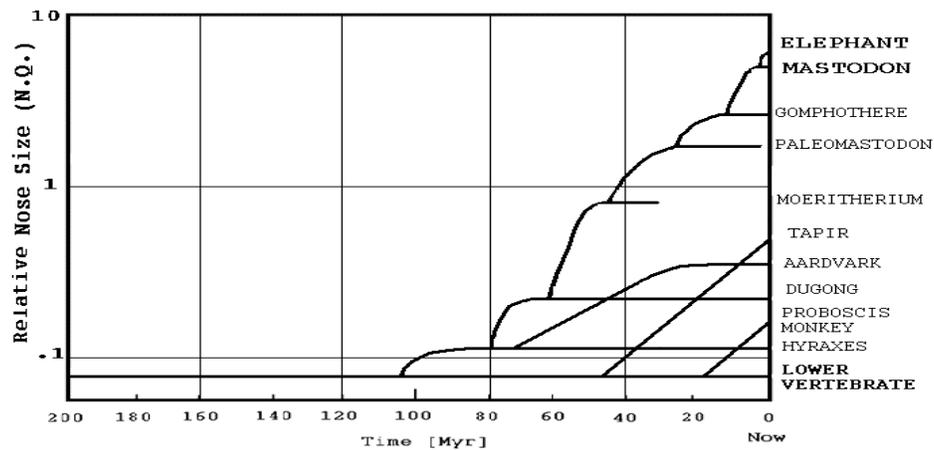

Figure 2. The Evolution of Relative Nose Size (= N.Q. =Nasalization Quotient, ratio of nose length to body length) Over the Past 200 Million Years. Notice the apparent trend in the data as, over time, N.Q. reaches its ultimate value in extant pachyderms. Notice also that once the direct lineage that led to elephants is ignored, most of the species do not have an increasing N.Q. A few do (tapirs, aardvarks, proboscis monkeys) and such exceptions are discussed in "Interpretational Problem # 3". This preliminary plot is meant to illustrate a point, and should not be taken as more than a crude representation of a specious trend in N.Q. that has been largely ignored and poorly quantified by paleontologists. See Poulakakis et al 2002 and Benton et al 2005 for more detail.

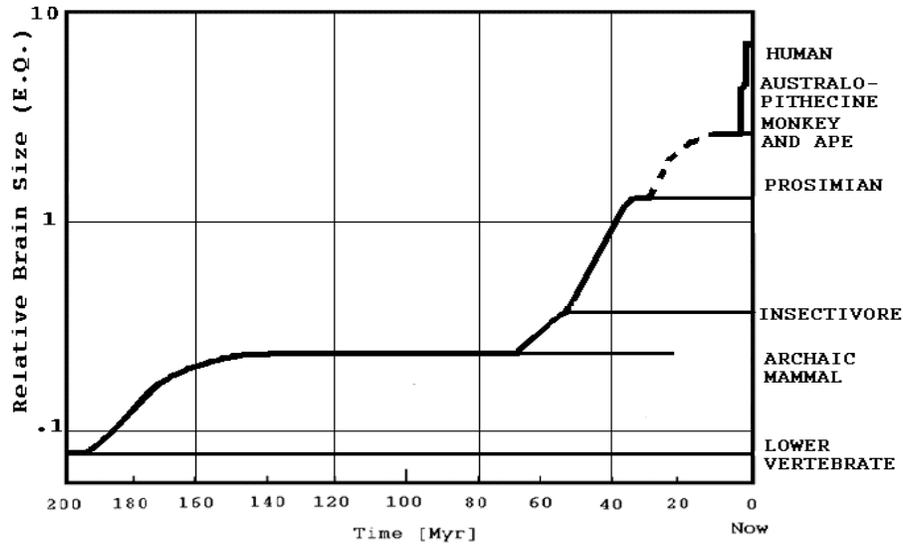

Figure 3. Same as Fig. 1 but we have removed the three lineages (dolphins, birds and ungulates/carnivores) with increasing E.Q. after diverging from our lineage. The lineage that leads to us provides no evidence of a generic trend towards increasing E.Q. because it has been selected to have that property. Notice that all of the increase in E.Q. is on the lineage that leads to humans. After diverging from us, all lineages shown here (lower vertebrates, archaic mammals, insectivores, prosimians, monkeys and apes) stayed at approximately the same E.Q. The lineages that remain flat represent the vast majority of all the species that evolved from the low E.Q. group shown in the lower left at 200 million years ago. The stasis of the E.Q. of the vast majority of species is evidence for an absence of selection pressure towards higher E.Q. The lineages whose E.Q. did increase after diverging from our lineage (dolphins, birds and ungulates/carnivores) are discussed in "Interpretational Problem # 3".

Another example: Humans have big brains with extremely small olfactory lobes. A similar analysis of the evolution of our olfactory lobe size would lead to the conclusion that the shrinkage of olfactory lobes is a generic trend. Such a conclusion would be misguided. By selecting an extreme outlying feature of an extant organism (whether big or small, short or long, hard or soft—anything as long as it is an outlier) the evolution of that feature in that organism's ancestors is very likely to show a monotonic trend that leads up to the outlier (Fig. 3).

### 5. Interpretational Problem # 2 Selection Bias: Non-Democratic Line Assignments

In Fig. 1 some lines represent one species (dolphin, humans) while other lines represent thousands of species (archaic mammals) or tens of thousands of species (lower invertebrates, see Dawkins 2005). If our goal is to fairly represent what has happened during evolution then each line should represent an independent cladistic group: order, class or genus – something democratic. However, having some lines for single species and other lines for thousands of species is a biased non-representative form of gerrymandering. If one is looking for a trend, one needs to consider all the data, not just the data that support the trend one wants to show. If the evolution of the E.Q.s of all species were plotted, some would go up, some would go down and the vast majority would stay the same. This variation is not shown in Fig. 1.

**6. Interpretational Problem # 3: Non-independence of "convergence" on high E.Q.**

Consider Fig. 1 again. After diverging from our lineage, the increasing E.Q. of birds, dolphins and ungulates/carnivores **does** seem to be evidence for the trend in the fossil record toward intelligence that Drake and Sagan were referring to. About three hundred and ten million years ago the last common ancestor of birds and humans had a small brain (E.Q. < 0.1). And 310 million years later birds and humans have bigger brains (E.Q. ~ 0.8 and E.Q. ~ 8 respectively). About eighty five million years ago the last common ancestor of dolphins, humans and ungulates/carnivores had a small brain (E.Q. ~ 0.2). And 85 million years later humans and dolphins have a big brain with E.Q. ~ 8 and ungulates/carnivores have an E.Q.~ 1. The E.Q. in all three lineages "independently" got bigger. Can we extrapolate this independent convergence on high E.Q. to the evolution of extraterrestrials?

Simon Conway-Morris (2003, 2005) has documented many cases of evolutionary convergence in evolution: – both marsupial and placental mammals converged on saber-toothed carnivores (thylacosmilids and placental cats). The ability to fly evolved in insects, pterosaurs (reptiles), birds and bats (mammals). Conway-Morris and other authors have cited N independent examples of the origin of the eye, where N is some largish but indeterminate number.

The common ancestor of these "independent" eye inventors did not have easily identifiable eyes, but it almost certainly did have proto-eyes of some sort. The supposed independence of these convergences is undermined by the ~3.5 billion years during which the creatures were biochemically and genetically identical, and during which they evolved their proto-eyes. They shared the same genes, the same genetic machinery of gene expression, regulation, inhibition and activation that controlled the genetic exploration of morpho-space and the ability to tinker with the structure of the head and the placement and framing of photoreceptors. Many shared the same head and brain. The supposedly independent evolvers of eyes, share the same basic biochemistry and photoreceptor proteins and the same plasticity (West-Eberhard 1989) that enabled and constrained the morphological, preservable features that are superficially different enough to be called "independent".

When considering convergence, a basic principle is often ignored: the extent of convergence cannot be larger than the extent of divergence from the common ancestor. With all terrestrial life having a common origin, one must first quantify the degree of divergence of two groups before one can discuss their convergence. For the species that converged on eyes, this divergence could only take place during the relatively brief fraction of their existence as independent organisms. Roughly speaking, and depending on the eyes, the organisms were independent for about 500 million years but shared a common ancestor for about 3500 million years. Thus they were independent divergers for only ~10%-20% of their existence (~500 Myr / 4000 Myr ~ 12%) and were identical for ~80%-90% of their existence. In other words, the common ancestor of the independent eyes had, during ~ 3500 million years, already evolved the complex biochemical pathways for photoreception. In many detailed and fundamental biochemical and genetic ways, the purported "independent" originators (although they were phylogenetically isolated) were working in the same workshop, with the same tools and the same materials with the same set of genetic regulatory skills.

Similarly, the common ancestor of dolphins and humans who lived ~ 85 million years ago had a head, a small brain and ~ 500 million year common history of regulatory genes that tinkered with the characteristics of that brain. It had the same, (or very similar) biochemical neural pathways and genetic plasticity and constraints that dolphins and humans are still endowed with. This 500 million year history produced a finite number of correlated non-independent ways to adapt to environmental challenges. In other words, there were a finite number of highly evolved toggle switches that could be successfully tinkered with. Among the two thousand species of Laurasiatheres that diverged from our lineage 85 million years ago with the dolphins and carnivores/ungulates, all had heads and all had a similarly constrained potential for modification. There were thousands of species, undergoing a variety of environmental challenges, and their ability to adapt to these challenges was provided by hundreds or thousands of shared toggle switches. We would expect some apparent convergences. What I am calling genetic "toggle switches" is more articulately and authoritatively described as "conserved core processes", "facilitated variation" and "invisible anatomy" by Kirschner and Gerhart (2005).

Similar arguments can be made about the increased nasalization quotients in tapirs, aardvarks and proboscis monkeys (Fig. 2). They too converged on large noses. Once noses were present and a common genetic tinkering apparatus had become available from a long shared history, nose sizes increased and decreased but mostly (like brains) just stayed the same size.

Three and a half billion years of identical evolution comes with much biological baggage and many constraints and it is these limited choices that are largely responsible for the apparent convergence on big brains. What Drake, Sagan and Conway-Morris have done is interpret correlated parallel moves in evolution as if they were unconstrained by shared evolution but highly constrained by a universal selection pressure towards intelligence that could be extrapolated to extraterrestrials. I am arguing just the opposite -- that the apparently independent evolution toward higher E.Q. is largely constrained by shared evolution with no evidence for some universal selection pressure towards intelligence. If this view is correct, we cannot extrapolate the trends toward higher E.Q. to the evolution of extraterrestrials. If the convergence of dolphins and humans on high E.Q. has much to do with the 3.5 Gyr of shared history (and I argue that it has everything to do with it) then we are not justified to extrapolate this convergence to other extraterrestrial life forms that did not share this history. Extraterrestrials are related to us in the sense that they may be carbon and water based – they may have polymerized the same monomers using amino acids to make proteins, nucleotides to make a genetic code, lipids to make fats and sugars to make polysaccharides. However, our "common ancestor" with extraterrestrials was probably pre-biotic and did not share a common limited set of genetic toggle switches that is responsible for the apparently independent convergences among terrestrial life forms.

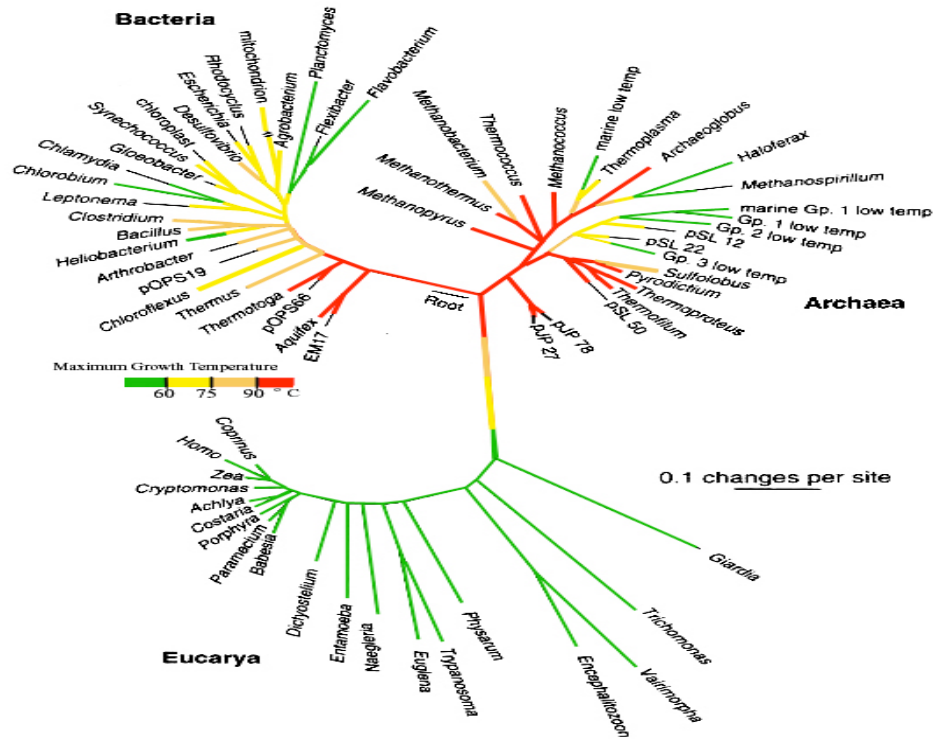

Figure 4. Phylogenic tree of terrestrial life based on the 16s subunit of ribosomal RNA. The last common ancestor of all life is at the center of the tree (labeled "Root"). The distance from the root to the end of each branch corresponds to the same amount of time – roughly 3.5 or 4.0 billion years. Because the ticking of the 16s molecular clock is not exactly uniform, the distances from the root to the ends of the branches are not the same length. Among the Eucarya in the lower left are the three twigs of complex multicellular life: Coprinus (representing fungi), Homo (humans, representing animals) and Zea ( corn, representing plants). The common ancestor of fungi, animals and plants lived ~1.5 billion years ago (Hedges et al 2004). Thus, the 200 million year time frame shown in Figs. 1 and 2 corresponds to the last ~2 mm of the twig labeled "Homo". Diagram from Lineweaver and Schwartzman (2004) based on Pace (1997). This RNA tree should be compared to Fig. 6.

We can use Fig. 4 to identify trends in the evolution of life, or convergences on some specific feature, whether it be E.Q., N.Q., olfactory lobe size or eyeballs. First we randomly select a few of the ~60 branches shown. Then we determine if two or more of them have independently evolved the feature of interest. For example, human-like intelligence probably depends on the existence of heads. Thus, we want to know if the tree of life shows any convergence towards heads. If heads were a convergent feature of evolution one would expect independent lineages to evolve heads. Our short twig on the lower left labeled "Homo" has heads, but heads are found in no other branch. Our two closest relatives, plants and fungi, do not seem to have any tendency toward evolving heads. The evolution of heads (encephalization) is therefore not a convergent feature of evolution. Heads are monophyletic and were once the possessions of only one quirky unique species that lived about six or seven hundred million years ago. Its ancestors, no doubt possessed some kind of proto-head related to neural crests and placodes (Wada 2001, Manzanares and Nieto 2003).

Drake (2006) stated that "[intelligence] is not a fluke that has occurred in some small sub-set of animal life." However, Fig. 4 shows that intelligence, heads, even all animal life or multicellular life, may well be a fluke that **is** a small sub-set of terrestrial life. One potential problem with this conclusion: It is possible that existing heads could have suppressed the emergence of subsequent heads. Such suppression would be difficult to establish.

## 7. Interpretational Problem #4: The Tenuous Link Between High E.Q. and Human-like Intelligence

About 600 million years ago, two kinds of metazoans, protostomes and deuterostomes, diverged from each other. Both evolved separately for ~600 million years and were very successful. Today there are about a million species of protostomes and about 600,000 species of deuterostomes (of which we are one). We consider ourselves to be the smartest deuterostome. The most intelligent protostome is probably the octopus. After 600 million years of independent evolution and despite their big brains, octopi do not seem to be on the verge of building radio telescopes. The dolphinoidea evolved a large E.Q. between ~60 million years ago and ~20 million years ago (Marino et al 2004). Thus, dolphins have had ~20 million years to build a radio telescope and have not done so. This strongly suggests that high E.Q. may be a necessary, but is not a sufficient condition for the construction of radio telescopes. Thus, even if there were a universal trend toward high E.Q., the link between high E.Q. and the ability to build a radio telescope is not clear. If you live underwater and have no hands, no matter how high your E.Q., you may not be able to build, or be interested in building, a radio telescope.

## 8. A Universal Intelligence Niche?

Life has been evolving on this planet for ~ 4 billion years. If the Planet of the Apes Hypothesis is correct and there is an intelligence niche that we have only recently occupied -- Who occupied it 2 billion years ago, or 1 billion years ago or 500 million years ago? Stromatilites? Algae? Jellyfish?

Sagan defines "the functional equivalent of humans" so narrowly (creatures able to build and operate radio telescopes) that only one species on Earth belongs to it – and then assumes that it is so broad that many aliens will fit into it (Fig. 5). He is postulating an imaginary group of species with only one species in it. Most biologists refuse to take the idea of such an imaginary group seriously. In studying the variety of life on this planet, they see that general groups with only one species in them are self-contradictions that do not exist – probably not a sound foundation upon which to build our hypotheses about extraterrestrial life.

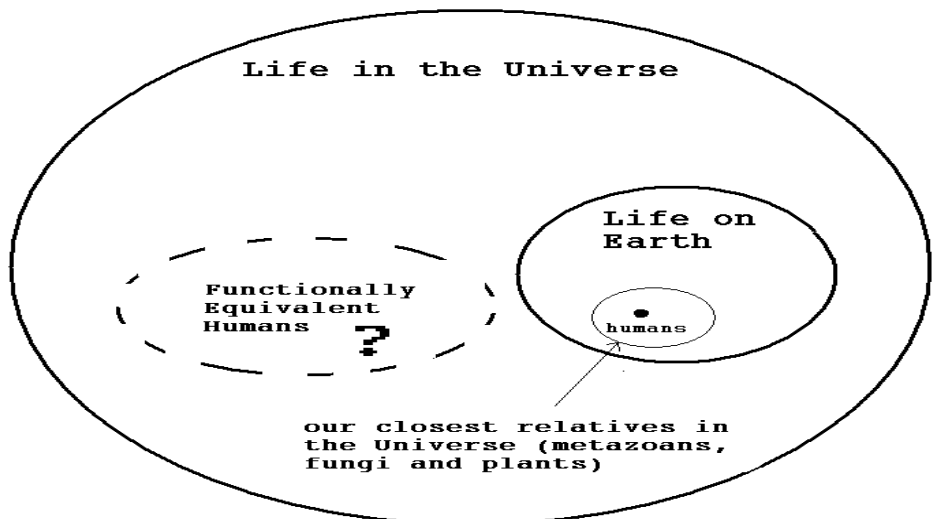

Figure 5  Consider the subsets of life in the Universe.  We know that humans (black dot) are a subset of Life on Earth and we know that metazoans, fungi and plants are our closest relatives on Earth (small circle around the black dot).  Sagan and others postulate an imaginary group of "functionally equivalent Humans" to which we (and some aliens) belong but none of our closest relatives do.

It seems unreasonable to define intelligence so narrowly that only Homo sapiens have it on Earth (among the ~100 million species that have ever lived)   and then imagine that the human-like intelligence niche is so generic that even life forms very different from ours (not sharing 3.5 billion years of evolution) would evolve into it.

   Any given species that has evolved on the Earth will have its closest relatives here on Earth.  Thus, if we consider humans to be unique and alone on Earth, then humans are a fortiori unique and alone in the Universe.  We are more closely related to the life forms with whom we have shared 3.5 billion years of common ancestors than we will be with any alien evolved independently on another planet.  Our closest relatives, genetically, physiologically and mentally are here on this Earth.

**9. Conclusion**

The search for extraterrestrial intelligence is a search for ourselves.  And therein lies its strength and weakness.  Knowing that we are searching for ourselves gives us the strength and motivation to explore and find our place in the Universe.  The weakness is the labyrinth of dead-ends created by our natural sense of self-importance and by our bias about what our place should be.  Figure 6 is an older tree than Fig. 4 and displays more obviously what we want to believe about ourselves.

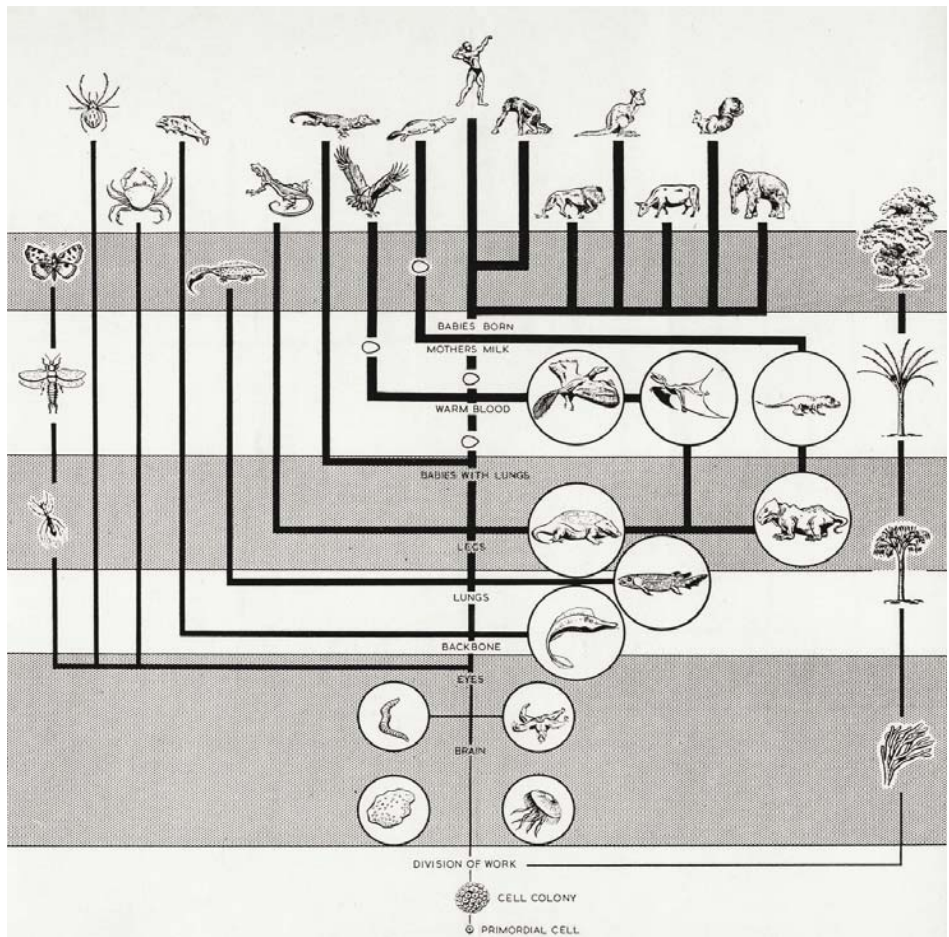

Figure 6. The Schwarzeneggerization of Life. A muscle-bound man stands as the end product of a linear progression—the Great Chain of Being –a ladder of life that leads to male Caucasian weight lifters. One can create such an apparent linear trend out of the crooked phylogenetic branch of **any** species. Looking back from any particular species we will find the evolution of the traits of that particular species. However, precisely because we can construct such a figure from the lineage of **any** species, such a construction should not be construed as a general linear trend applicable to all life. The simple appeal of this figure is a good example of how easy it is to believe that the important events and the major transitions in evolution that led to us, are important events for all organisms (Smith and Szathmary 1995). The problems with this view are detailed in Gould (1989). The prevalence and recurrence of this mistaken interpretation of evolution needs to be avoided as we try to use terrestrial evolution to give us hints about the evolution of extraterrestrial life. Figure from Gatland and Dempster (1957). This homo-centric tree should be compared with Fig. 4.

If human-like intelligence were so useful, we should see many independent examples of it in biology, and we could cite many creatures who had involved on independent continents to inhabit the "intelligence niche". But we can't. Human-like intelligence seems to be what its name implies -- species specific.

I have argued that the fossil record strongly suggests that human-like intelligence is not a convergent feature of evolution. The evidence is indirect and suggestive, but it is, I think, the best we have. Despite this evidence, I am a strong supporter of SETI -- because I may be wrong about how the evidence is best interpreted, and because SETI is relatively cheap science. SETI is the exploration of new parameter space with new instruments – a proven recipe for scientific discovery. However, we do not need to misinterpret the fossil record to justify this inspiring research.